\documentclass[review]{elsarticle}

\usepackage{lineno,hyperref}
\usepackage{subcaption}
\usepackage{amsmath}
\usepackage{cleveref}
\usepackage{bbm}

\modulolinenumbers[5]

\journal{Geoderma}

\newcommand{\nc}{\ensuremath{n_{\text{classes}}}}
\newcommand{\ns}{\ensuremath{n_{\text{samples}}}}
\newcommand{\indy}[1]{\ensuremath{  \operatorname{\mathbbm{1}}\left(#1 \right) }}

\begin{document}

\begin{frontmatter}

\title{Optimal soil sampling design based on the maxvol algorithm}


\author[mymainaddress]{Anna Petrovskaia\corref{cor1}}
\ead{anna.petrovskaia@skoltech.ru}
\author[mymainaddress]{Gleb Ryzhakov}
\author[mymainaddress,mysecondaryaddress]{Ivan Oseledets}

\cortext[cor1]{Corresponding author}

\address[mymainaddress]{Skolkovo Institute of Science and Technology, Bolshoy Boulevard 30, bld. 1, Moscow, Russia 121205\\
\url{www.skoltech.ru}}
\address[mysecondaryaddress]{Marchuk Institute of Numerical Mathematics, RAS, Gubkin st 8, Moscow, Russia 119333\\
\url{www.inm.ras.ru}}

\begin{abstract}
Spatial soil sampling is an integral part of a soil survey aimed at creating a soil map. We propose considering the soil sampling procedure as a task of optimal design. In practical terms, optimal experiments can reduce experimentation costs, as they allow the researcher to obtain one optimal set of points. We present a sampling design, based on the fundamental idea of selecting sample locations by performing an optimal design method called the maxvol algorithm. It is shown that the maxvol-base algorithm has a high potential for practical usage. Our method outperforms popular sampling methods in soil taxa prediction based on topographical features of the site and deals with massive agricultural datasets in a reasonable time.
\end{abstract}


\end{frontmatter}


\section{Introduction}

Spatial soil sampling is an integral part of a soil survey aimed at creating a soil map. This step dramatically effects the quality and accuracy of a map, as well as the cost of a survey \citep{hengl2003soil}. 

Ideally, an optimal layout of soil sampling points should contain as few points as possible, and at the same time, it should capture the variability of soil cover suﬃcient for creating a precise soil map.

Since precision agriculture usage and environmental management saw a rapid growth in recent years, various soil sampling approaches have been proposed \citep{brus2007optimization, zhu2008purposive, fitzgerald2006directed, clifford2014pragmatic,liess2015sampling,musafer2016optimal, nawar2018optimal, nketia2019new, aktas2019landslide, wadoux2019sampling,castaldi2019sampling, liess2020interface,ma2020comparison, yang2020evaluation}. Diﬀerent modifications of Latin Hypercube sampling, simple random sampling, and grid sampling are commonly investigated ways to achieve an optimal sampling scheme. Conditioned Latin Hypercube (cLHS) \citep{minasny2006conditioned} is among the most widely used algorithms. 

 Usually, soil sampling strategies are divided into two fundamentally diﬀerent groups: design-based and model-based approaches  \citep{de2006sampling, brus1997random}. In a design-based approach, also called "classical statistical" sampling, sampling locations are selected by probability sampling, and the statistical inference is based on the sampling design. In model-based sampling, also called the "geostatistical" approach, a model for the spatial variation is introduced \citep{brus2012hybrid}. 
 
It is discernible that the prevailing trend approaches the problem of spatial soil sampling by sampling from a probability distribution. However, instead of looking at the problem in a common way, it is natural to consider the soil sampling procedure as a task of optimal design. Optimal design is a class of experimental designs that are straight optimisations based on a chosen optimality criterion \citep{OptDesign, natrella2010nist}.
 The usage of optimal designs in statistical models’ estimation allow parameters to be evaluated without bias by minimising the overall variance. A non-optimal design requires a higher number of experimental runs to determine the same precision settings as an optimal design. In practical terms, optimal experiments can reduce experimentation costs, as they allow one to obtain one optimal set of points. D-optimal design methods for sampling were mentioned by \cite{brus2007optimization} as possible methods for model-based mapping. However, there is still no commonly used soil sampling design based on optimal design theory.

In this paper, we present a sampling design, the fundamental idea of which is to select sample locations by performing a D-optimal design method called the maxvol algorithm. The advantage of the proposed approach is that it is adaptive to input data. As predictors, we selected variables that in view of previous findings, seemed likely to play a role in soil heterogeneity on a large scale \citep{jana2012topographic, jana2012topography,kozlov2017soil, lozbenev2021incorporating}. We propose transforming a digital elevation model (DEM) to a matrix of features of size m pixels by n features by computing topographical features in each point. The maxvol algorithm is then used to select locations on a study site by choosing a row in the feature matrix that corresponds to a particular pixel with the most significant dissimilarities in topographical features.

This paper’s main contribution is a new sampling design, which shows better results in predicting soil variability with a considerably smaller dispersion in prediction error than the existing approaches for spatial soil sampling. This paper outlines the theoretical background of the maxvol algorithm; then, it presents the methodology of sampling design based on the maxvol algorithm. The proposed sampling design is then tested in three real cases diﬀering in field data availability and compared to the popular sampling designs — conditional Latin Hypercube, simple random sampling, and Kennard–Stone \citep{kennard1969computer}.

\section{Materials and methods}

\subsection{The maxvol algorithm}


We propose a sampling design based on an optimal design approach. As the primary purpose of sampling design is to provide the most comprehensive information about the site, we can consider it as a procedure of finding the knowledge of a suﬃciently well-conditioned submatrix in a large scale matrix. In this case, the best submatrix will be a sequence of rows, each of which is devoted to a particular place (pixel) on a study site. A natural fit for this problem is an optimal design algorithm, which has been used for a variety of tasks, such as recommender systems \citep{liu2011wisdom}, wireless communication \citep{wang2010global} and function approximation \citep{Ryzhakov2018}.

Optimal design methods allow finding a good submatrix by optimising some statistical criterion. One way to measure the submatrix quality is its determinant \citep{goreinov2010find}. We propose to use the maxvol algorithm, which finds a submatrix with the largest possible determinant, or in other words, the maxvol uses the D-optimality criterion for finding a good submatrix. Geometrically, the determinant of a matrix can be viewed as the volume scaling factor of the linear transformation described by the matrix \citep{de1995d}.This is also the sign volume of the n-dimensional parallelepiped spanned by the matrix’s column or row vectors. Thus, returning to the idea of comprehensive information, the maxvol algorithm maximises the volume of the parallelepiped to construct a submatrix with rows that describe the distribution of predictors in the most precise way. 

In our work, we follow the definitions from the work of \citep{goreinov2010find}. The modulus of a determinant of a square matrix is referred to as its volume. By definition, the determinant is calculated from a square matrix. However, in our work, we constructed a maximum volume submatrix, not for a square matrix, but a tall matrix, i.e., the matrix where only one dimension is large. 
By the volume of a square matrix  ($\operatorname{vol}_{1}(A)$) and tall rectangular matrix  ($\operatorname{vol}_{2}(A)$), we mean the following quantities:
\begin{equation}
\operatorname{vol}_{1}(A)=|\operatorname{det}(A)|,
\;\;
\text {vol}_{2}(A)=\sqrt{\max \left(\operatorname{det}\left(A^{H} A\right), \operatorname{det}\left(A A^{H}\right)\right)},
\end{equation}
where $A^H$ defines the conjugate transpose of a matrix $A$, $ \operatorname{det}(A)$ defines determinant of matrix $A$.

By definition, the maxvol algorithm is a greedy iterative algorithm, which swaps rows to maximise the volume of a square submatrix \citep{goreinov2010find}.Strictly speaking, for con-structing sampling design, we used a modified version of the maxvol algorithm, called rec\_maxvol \citep{mikhalev2018rectangular}.The rect\_maxvol allows the researcher to select rectangular submatrix $\hat{A}$
from the given matrix~$A$, which in practice means that our sampling algorithm can choose the required number of sampling points. In contrast to rect\_maxvol, plain maxvol could select the number of points only equal to the number of features. For simplicity in this work, we refer to modified rect\_maxvol as maxvol.

\begin{figure}[!h]
\center
\includegraphics [scale=0.45]{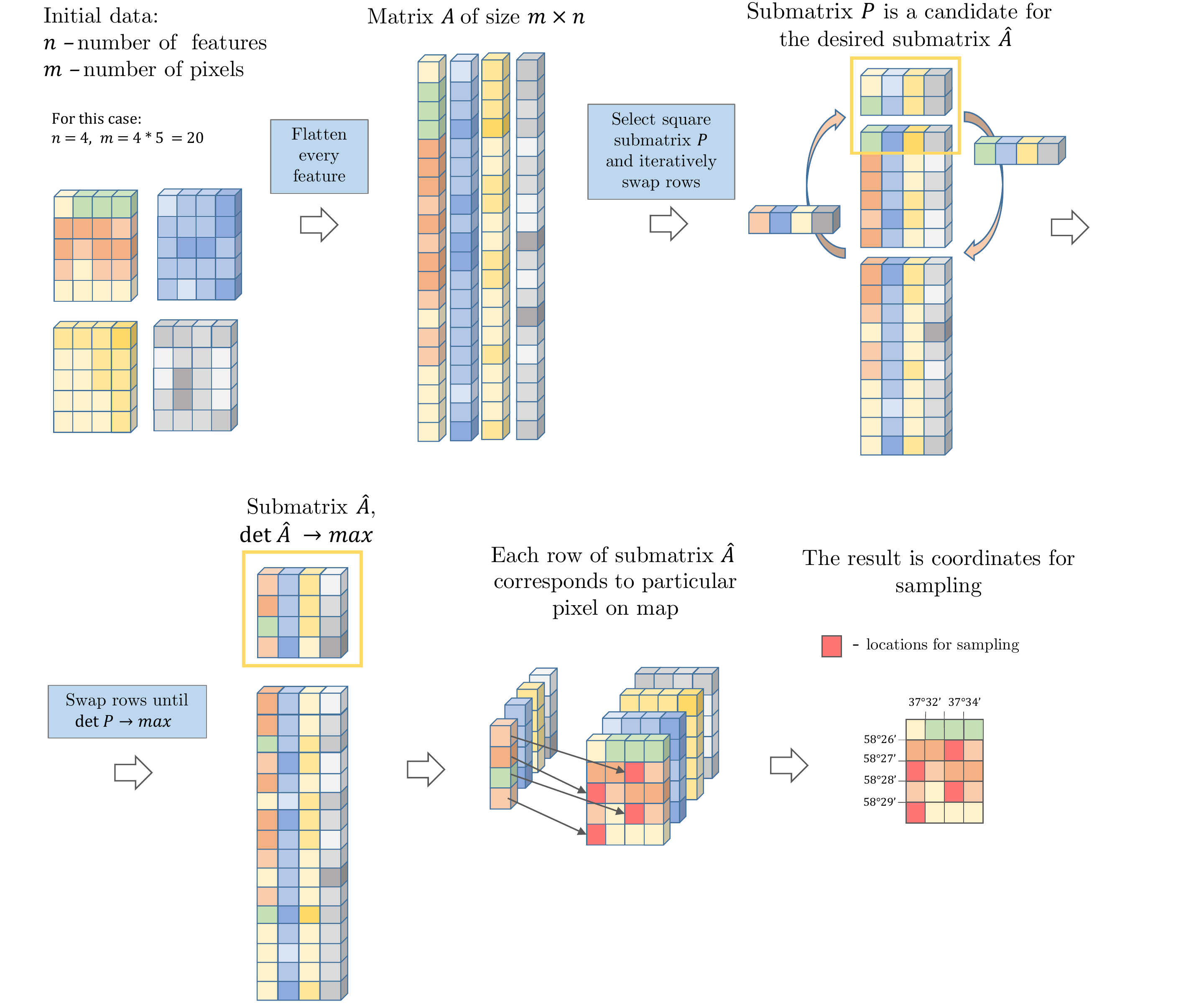}
\caption{Scheme of the application of the {\bf{maxvol}} algorithm to spatial sampling}
\label{maxvol} 
\end{figure}

The general scheme of sampling design is presented in  \cref{maxvol}. The initial data for the sampling algorithm are particular features of the study site of similar shapes. Firstly, we construct a matrix $A$ by flattening every feature. Therefore, the matrix $A$ has a size $ m \times n$, where $n$ is a number of features, and $m$ is a number of pixels in the initial image. In other words, every row of matrix $A$ is the set of feature values in a certain pixel of the initial map.  After that, the iterative procedure of finding the best submatrix $\hat{A}$ begins. In the beginning, we select the top rows of matrix $A$, defined as submatrix $P$. Submatrix $P$ is a candidate for being the desired submatrix $\hat{A}$. The rows are swapped between submatrix $P$ and the other rows of matrix $A$ until the determinant of $P$ reaches the maximum possible value. A precise description of the procedure for finding submatrix $\hat{A}$ can be found in the work of \citep{mikhalev2018rectangular}. When $P$ has the maximum possible value of determinant, it is assumed that we found submatrix $\hat{A}$. As each row of $\hat{A}$ corresponds to a particular pixel on the initial map, the result of the algorithm is locations on the study site being considered.

\begin{figure}[!ht]
  \center
  \includegraphics [scale=0.6]{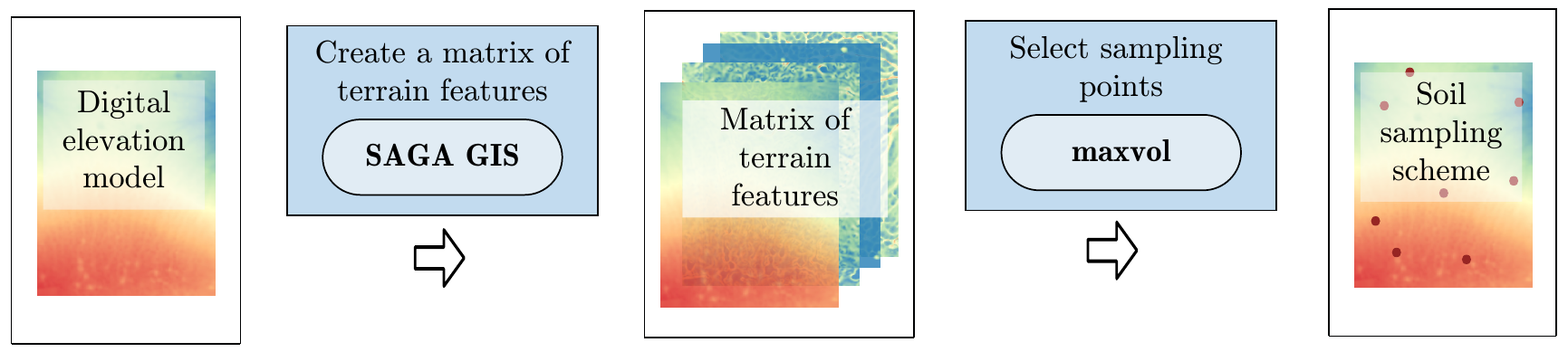}
   \caption{General scheme of the methodology}
   \label{gen_scheme} 
\end{figure}

In  our application, we used a matrix of land-surface parameters as a large matrix $A$ for approximation.  Thus, the workflow of our sampling design consists of a two-step procedure that includes creating an information matrix and selecting a subset of points using the {\bf{maxvol}} sampling algorithm ( \cref{gen_scheme}). 

\subsection{Distance constraint} \label{distance}
We adjusted our algorithm with a heuristic to make it more useful in real soil mapping practice. This heuristic lets a user restrict the minimum distance between sampling points. As the maxvol algorithm is allowed to select every pixel of the site, significant diﬀerences in topographical features may cause several nearby pixels to be chosen as sampling points. It is a known issue, mentioned by \cite{brus2007optimization},that optimisation in the feature space may lead to strong spatial clustering of sample locations and one needs to find a compromise between spreading in feature space and in geographic space. To find a compromise, we propose to use an option that restricts the minimum distance between sampling points. Distances between points are calculated based on the coordinates of pixels. 

This heuristic is placed into the sampling algorithm. The idea is to zeroise the rows selected as candidates for a submatrix $\hat{A}$ if the corresponding pixel on the initial map is closer to the previously selected points than some fixed~$\epsilon$. This value is defined manually with regards to the size of study site being considered.
The distance constraint is applying to every new pixel that is a candidate for sampling and has the following form: 
\begin{equation}
    (x_{\text{new}}-x_i)^2 + (y_{\text{new}}-y_i)^2<\epsilon^2, 
\end{equation}
where $x_{\text{new}}$ and $y_{\text{new}}$ are coordinates $x$ and $y$ of the point that is a candidate for sampling, $x_i$ and~$y_i$ are coordinates of the points that have already been selected.
If this condition is met, the candidate point is discarded. 

The choice of the distance constraint’s value requires further deep research. In this paper, the distance constraint was set manually based on the degree of terrain ruggedness on the sites: the more rough the terrain, the smaller the value of distance constraint. 

\subsection{Investigated areas}

\begin{figure}[!ht]
  \center
  \includegraphics [scale=0.08]{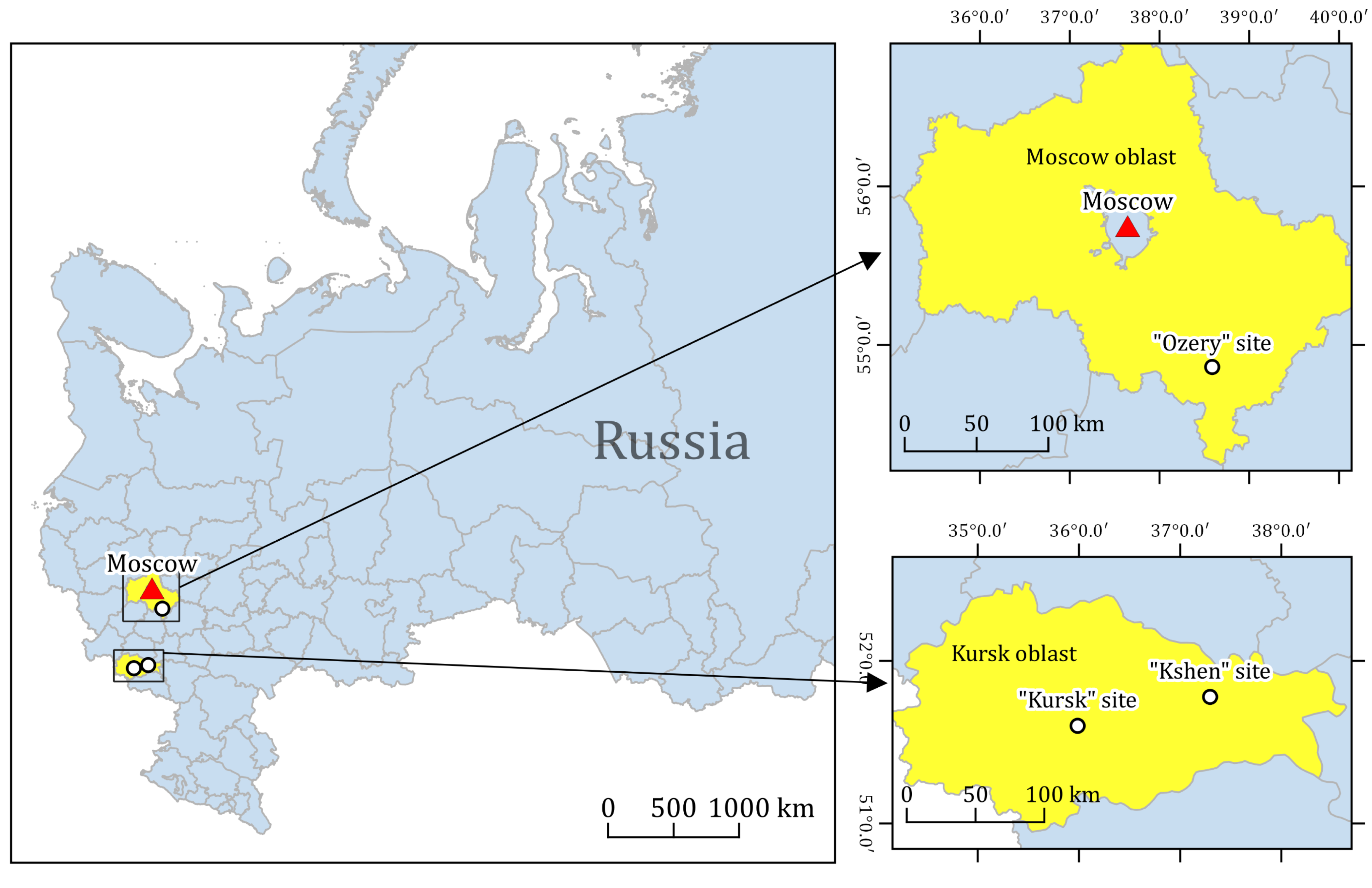}
   \caption{Location of study sites}
   \label{sites} 
\end{figure}

To verify the applicability to the diﬀerent cases of field data which are available, the developed algorithm was applied to three field landscapes. One of them is presented by a natural reserve, where several research institutes conducted long-term studies. The other two are cultivated agricultural fields that vary in shape and size and diﬀer in quality and quantity of available information. All three study sites are located in European Russia, ~\cref{sites}, Table~\cref{sites:table}.

The Central Black Earth Biosphere Reserve, where the first investigated area is located, is a collection of selected sites of black soil prairie in the southwestern part of the Central Uplands within the middle of the forest-steppe zone, which are protected for scientific study. The investigated site ("Kursk" site) comprises an area of approximately 0.4 $km^2$ of virgin forest-steppe. The data used for the testing of sampling algorithms was collected in 2013--2016 during the detailed mapping of the soil cover. 

The source materials for the site include a digital elevation model (DEM) and soil data. DEM has a resolution of 2.5 m, created with a  Global Navigation Satellite System (GNSS) survey. The size of DEM is 285 $\times$ 217 pixels, with a pixel size relative to 2.5 meters. Soil data on the site consists of 157 soil profiles, including information about various qualitative soil properties and the depth of carbonate occurrence. The soil points were located considering a specific hollow microrelief at the site.
It was revealed that four soil taxa present soil cover of the study area according to classification and diagnostics of soils of the USSR \citep{egorov1987classification}: chernozems typical (Haplic Chernozem according to WRB \citep{karklins2015world}),  chernozems typical calcareous with bioturbations (Haplic Chernozem), chernozems leached (Luvic Chernozem and Luvic Chernic Phaeozem),  and leached meadow chernozem (Luvic Stagnic Chernic Phaeozem) \citep{lozbenev2019digital}. 

One of the considered agricultural sites ("Kshen" site) occupies a cultivated field near Kshen, Kursk oblast. The area of this site amounts to 1.21 $km^2$.  Crop rotation is practiced on the field; consequently, crop on the field varies from year to year. The soil cover on the site was explored in 2017 when the field was covered by winter wheat. An agricultural company which owned the field provides us a yield map of the 2017 year. 
During the investigation, a  DEM for the site was created using an unmanned aerial vehicle (UAV). 
The soil cover of this site is presented by chernozem leached. There is no soil map in this dataset.

The second agricultural site lies in the Moscow region near Ozery ("Ozery" site). This site includes 7 cultivated agricultural fields used for growing vegetables, varying in a square from 0.95$km^2$ to 0.186 $km^2$. The "Ozery" site's dataset contains only a DEM which covers all 7 agricultural fields with a resolution of 5 meters per pixel. This dataset has no information about soil cover and yields, which is the most typical case in real practice, as placing sampling points is usually the first step of the survey.

\begin{table}
\centering
\caption{Data available for the study sites}
\label{sites:table}
\begin{tabular}{|p{0.2cm}|p{3.5cm}|p{2.5cm}|p{4.3cm}|}
\hline
 & \bf Location & \bf Description & \bf Materials\\
\hline
1 & Kursk oblast, Central Black Earth Biosphere Reserve  & Virgin forest-steppe & DEM (285*217 pixels, resolution - 2.5 m), 157 soil descriptions and soil map \\
\hline
2 & Kursk oblast, Kshen & Cultivated field & DEM (5108*9164 pixels, resolution - 5 m), yield map \\
\hline
3 & Moscow oblast, Ozery & Cultivated field & DEM (1991*2288 pixels, resolution - 5 m) \\
\hline

\end{tabular}
\end{table}


\subsection{Topographical features}
 The maxvol algorithm uses terrain features as predictors, based on the assumption that morphometry is one of the main contributors to soil variability on a large scale.In the previous studies, it was revealed that on "Kursk" site, the main factor in soil and plant cover diﬀerentiation is the redistribution of soil moisture along the microrelief \citep{kozlov2017soil,lozbenev2019digital}.
Additionally, elevation data is convenient to use in the beginning of a field survey because it can be easily obtained with an unmanned aerial vehicle. Still, the input to the maxvol could be other environmental gridded information depending on the availability of auxiliary data and dominating soil forming factors.

Five morphometric land-surface parameters were derived from the digital elevation model (DEM). The following morphometric parameters were obtained by SAGA GIS software \citep{saga}: aspect, closed depressions, slope, and topographic wetness index.
Also, using the SIMulated Water Erosion model of the Grass GIS software \citep{GRASS_GIS_software}, overland flow hydrologic simulation was calculated.

The result is images of the same size and resolution as the DEM, consisting of pixels with the values of particular parameters. All the values were normalised. For better points placement, morphometric parameters were supplemented with two coordinate layers, consisting of latitude and longitude values of every pixel. 

\begin{figure}[!ht]
  \center
  \includegraphics [scale=0.45]{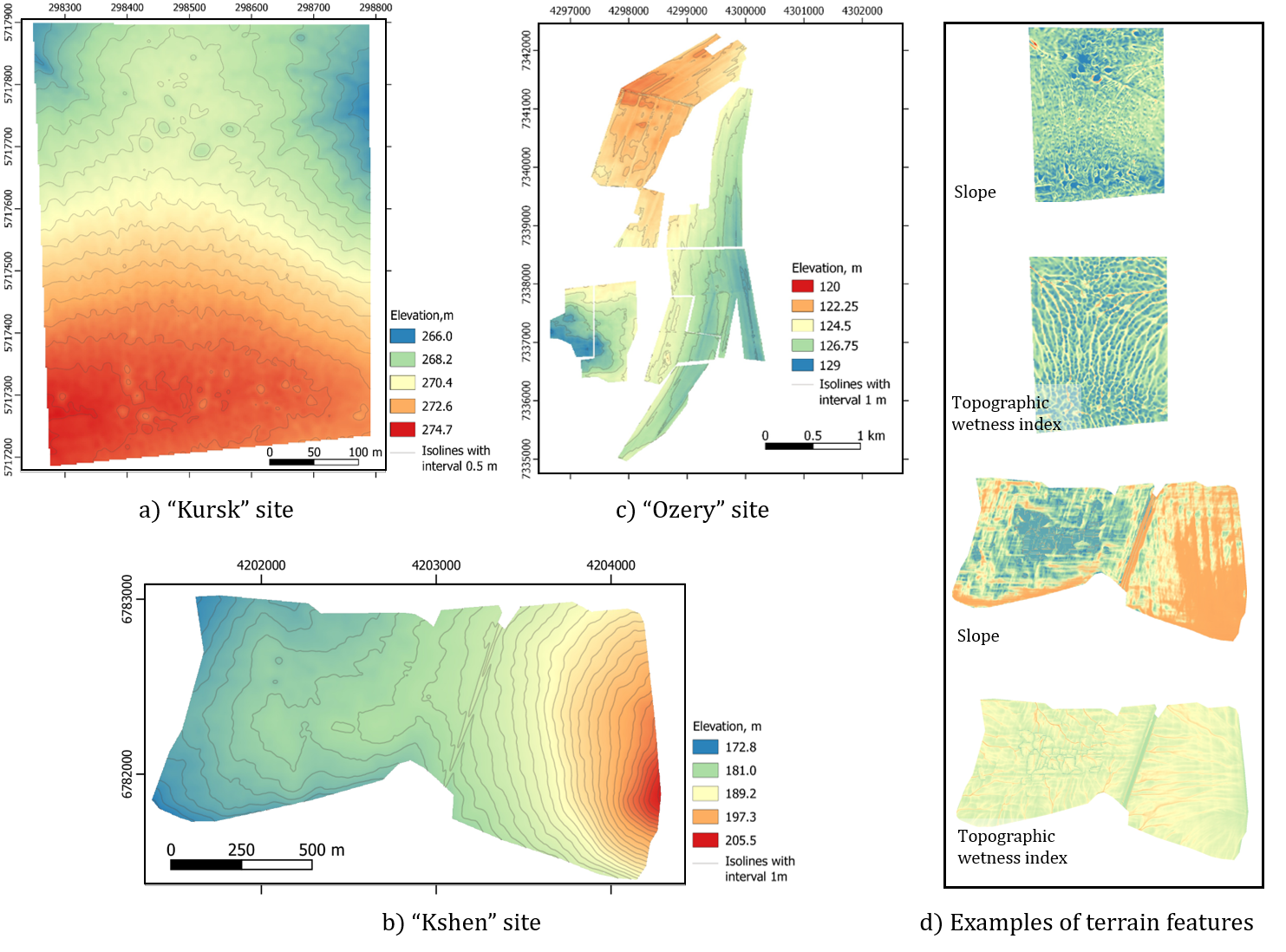}
   \caption{Digital elevation models of the sites (a, b, c) and the examples of terrain features (d)}
   \label{fig:proc} 
\end{figure}

\subsection{Evaluation methodology}\label{eval}

There are several ways to evaluate the eﬃcacy of the maxvol sampling algorithm according to the presence of information about soil cover on the site. We will consider two cases. In the first case the user of the algorithm has suﬃcient soil data to describe the whole distribution of soil types or properties on the siteIn this case the quality of the sample scheme produced by the algorithm could be measured by comparing the prediction of soil distribution based on the initial information and the prediction obtained by using soil information only in the points chosen by the algorithm.

The "Kursk" site is a case where there existed precise soil information before conducting soil sampling. Accordingly, to validate the accuracy of the algorithm in producing the soil sampling scheme, a digital soil map of the "Kursk" site was made. It was made based on the layout of 157 soil profiles using Naive Bayes classification. Accuracy estimation on the site was carried out by comparing the prediction of soil distribution based on the initial information (157 soil descriptions) and the predictions obtained using soil information only in the points chosen by the algorithm. The most common technique to estimate an algorithm's success quantitatively is the accuracy score, which is the fraction of correct predictions.
If $\hat{y}_{i}$ is the predicted value of the $i$-th sample and $y_{i}$ is the corresponding true value, then the fraction of correct predictions over
$n_{\text {samples}}$ is defined as 
\[
\operatorname{accuracy}(y,\, \hat{y})=\frac1\ns
\sum_{i=1}^{\ns} \indy{\hat{y}_{i}=y_{i}}
\]
where $\indy x$ is the indicator function of a boolean argument.

However, the classes in the data are imbalanced, so additionally we need to compare the performance of the algorithms with balanced accuracy. The original formula was adjusted by incorporating weights which determine the number of pixels in a class.

Let  $\{c_i\}$ for $i=1..\nc$ be a set of all possible values of $y_i$, 
where $\nc$ is the cardinality of this set.
Thus, for every index~$j=1..\ns$ there exists a unique index~$i$ such that $y_j=c_i$.
We define balanced accuracy as
\begin{equation*}
    \operatorname{balanced-accuracy} (y,\, \hat y)=
    \sum_{i=1}^{\ns} \indy{\hat y_i=y_i}
    \omega_i
\end{equation*}
with weights $\omega_i$ in the following form
\begin{equation*}
\omega_i=
\frac1\nc
\sum_{j=1}^{\nc}
\frac{ \indy{y_i=c_j} }{ \sum_{k=1}^{\ns} \indy{y_k=c_j} }
.
\end{equation*}
It should be noticed that we can simplify the expression of $\omega_i$ to the form
\begin{equation*}
\omega_i=
\frac1{\nc\cdot \sum_{k=1}^{\ns} \indy{y_k=y_i} }.
\end{equation*}

As the sampling procedure is the first step in a soil survey, usually there is no precise information about soil cover on a site. In this case we propose to use remote sensing information presenting the density of vegetation for accuracy estimation. 
Since vegetation distribution within a field depends on soil heterogeneity to a large extent  \citep{einsmann1999nutrient,hutchings2003toward, patzold2008soil}, the sampling scheme should capture the variability of vegetation density.

Source data on the "Kshen" and "Ozery" sites has no information about soil cover, which is the usual situation while conducting agricultural tasks for revealing soil cover properties and structure. For estimation of the sampling scheme’s quality on the "Kshen" site, the yield map was used. On the "Ozery" site, we evaluate the maxvol performance by the analysis of point allocation on the various landforms.

\section{Results}
This section demonstrates the applicability of the {\bf{maxvol}} algorithm to real field data. The algorithm is implemented in Python and is available online as \texttt{maxvolpy}\footnote{https://github.com/c-f-h/maxvolpy} library.  We compared the developed algorithm's performance with three sampling algorithms: cLHS, Kennard--Stone, and random sampling for 3 sites presented by virgin and cultivated fields. All the experiments provided in this paper are available at Github repository\footnote{https://github.com/petrovskaia/maxvol-for-soil}.

\subsection{Experiment with virgin field}

\subsubsection{Maxvol perfomance}

Sampling schemes obtained by the {\bf{maxvol}} algorithm based on topographical features are presented in  \cref{fig:points_graph}.  Fig.\ref{fig:sub1} represents 16 sampling points chosen by the plain {\bf{maxvol}} algorithm without penalties. The points’ positions follow the major irregularities of a land surface on the site. Points are placed at the site’s main landforms — the elevated part of the site, hollows in the upper left and right corners, and the closed depression in the centre of the slope. Several sampling points are provided at the footslope. Two groups of points were located quite close to each other: in the left hollow (yellow square,  \cref{fig:sub1}) and the closed depression (red square,  \cref{fig:sub1}). 

The right figure shows the same number of points allocated on the site utilising the maxvol algorithm adjusted by 'distance' option.  This option controls the distance between points (precisely described in \cref{distance}). The value of distance constraint is 0.1, so the smallest possible distance between the points is approximately 54 meters. In this case, the points layout captures land-forms more diversely. The hollows on the left and the right are provided by points near the bottom of the hollow and at the beginning of the hollow's slope. Also, a point was added to the second depression in the centre of the slope, and two points were placed forther up the slope.

\subsubsection{Comparison with other methods}

The maxvol algorithm’s performance was compared to several popular sampling methods: cLHS, Random Sampling, and KennardStone. The evaluation of sampling schemes was conducted according to the methodology described in\cref{eval}. Every algorithm was tested with the number of sampling points varying from 7 to 27.  Random sampling and cLHS, to some extent, are randomised algorithms and have a dispersion in the output, so we tested them 1000 times for every number of points and show an interval of 5\% and 95\% percentiles and a mean value. Every result of the cLHS algorithm was obtained with the number of iterations equal to 10000. 

\begin{figure}
  \centering
  \begin{subfigure}{.45\linewidth}
  \centering
  \includegraphics[width=\linewidth]{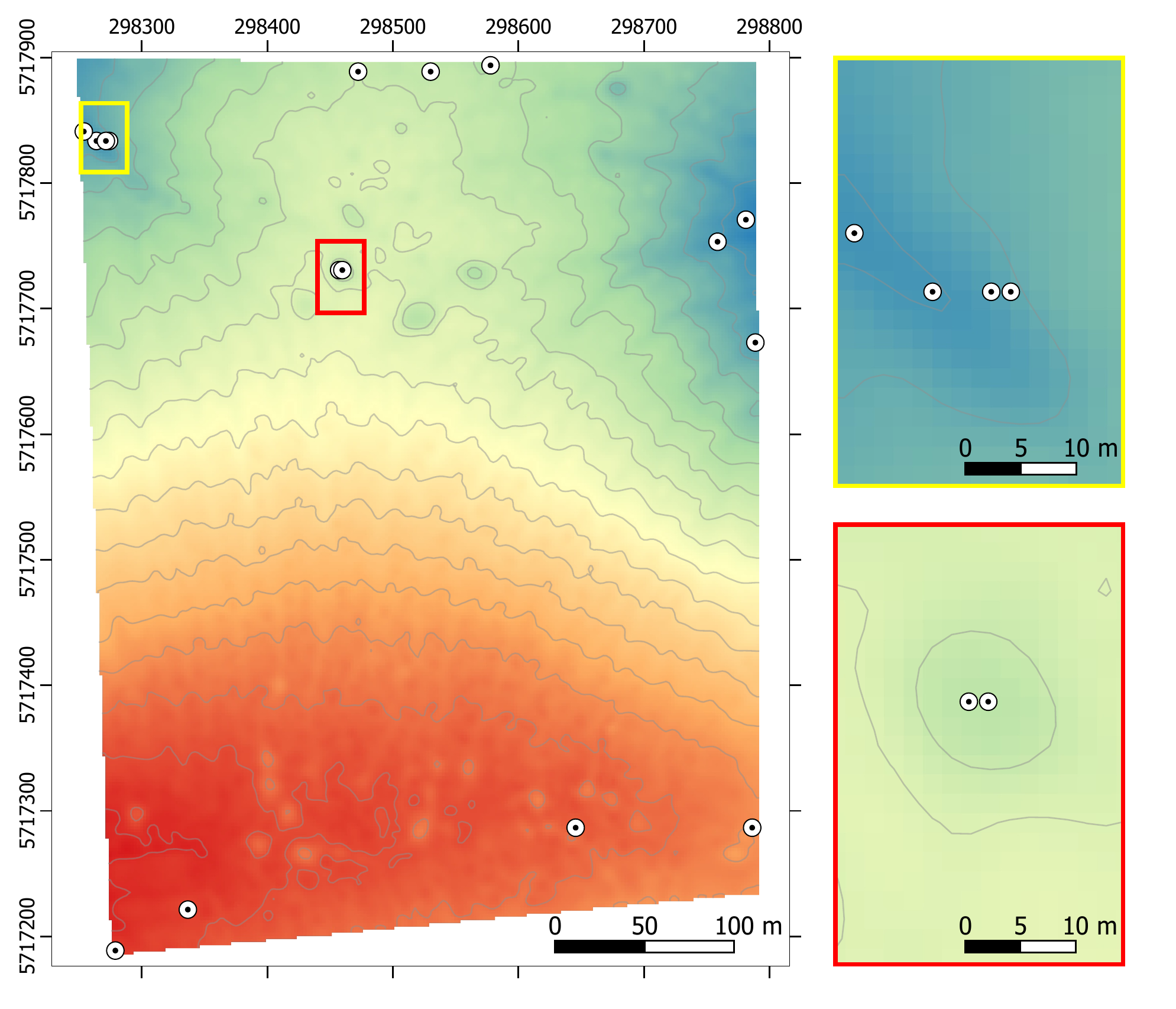}
  \caption{Soil sampling scheme produced by the {\bf{maxvol}} without distance constraint}
  \label{fig:sub1}
\end{subfigure}\hspace{2mm}%
\begin{subfigure}{.45\linewidth}
  \centering
  \includegraphics[width=\linewidth]{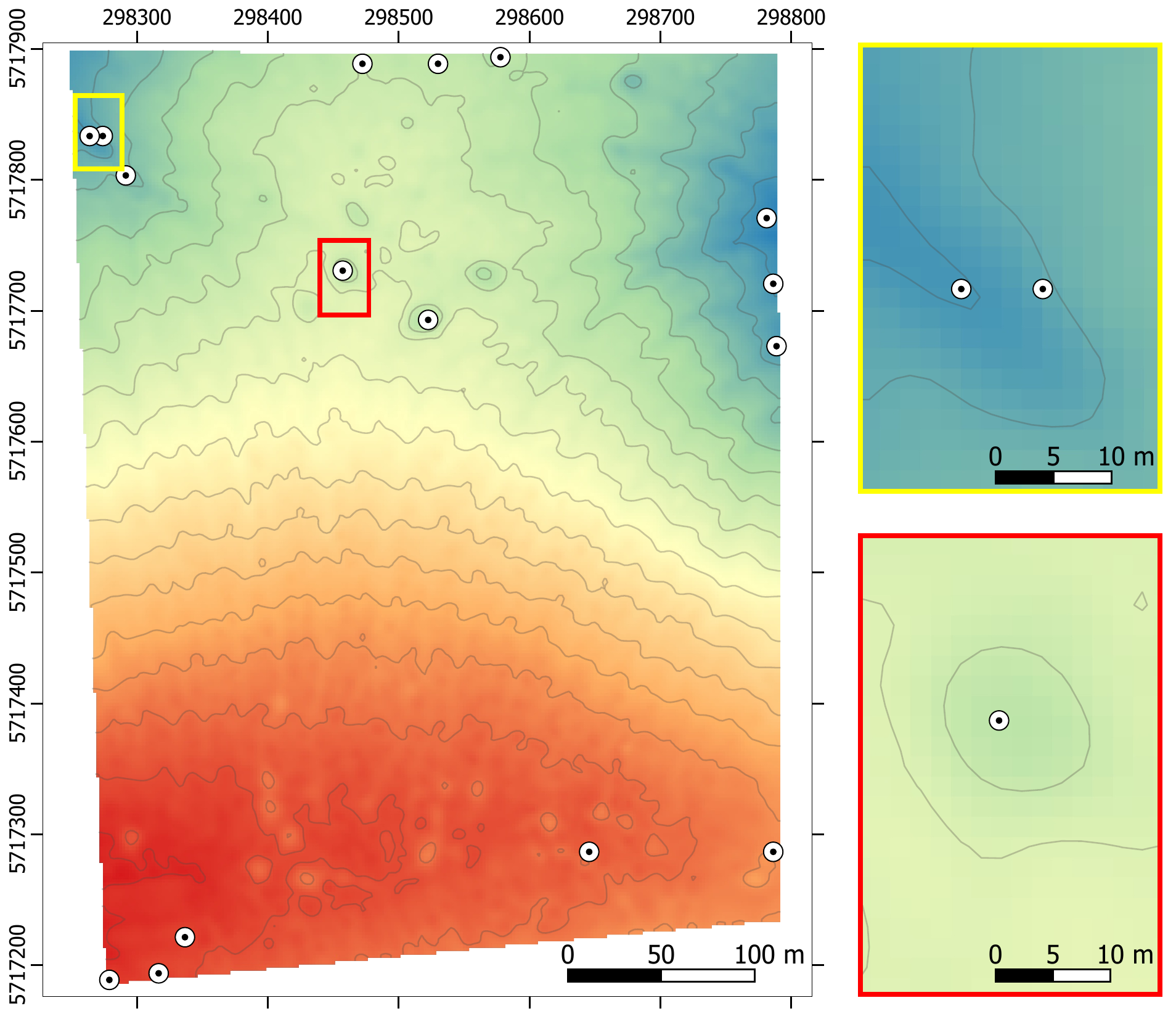}
  \caption{Soil sampling scheme produced by the {\bf{maxvol}} with distance constraint ($\epsilon$=0.1)}
  \label{fig:sub2}
\end{subfigure}
   \caption{Sampling schemes produced by the {\bf{maxvol}} on "Kursk" site}
   \label{fig:points_graph} 
\end{figure}

\begin{figure}
\centering
\begin{subfigure}{.5\linewidth}
  \centering
  \includegraphics[width=\linewidth]{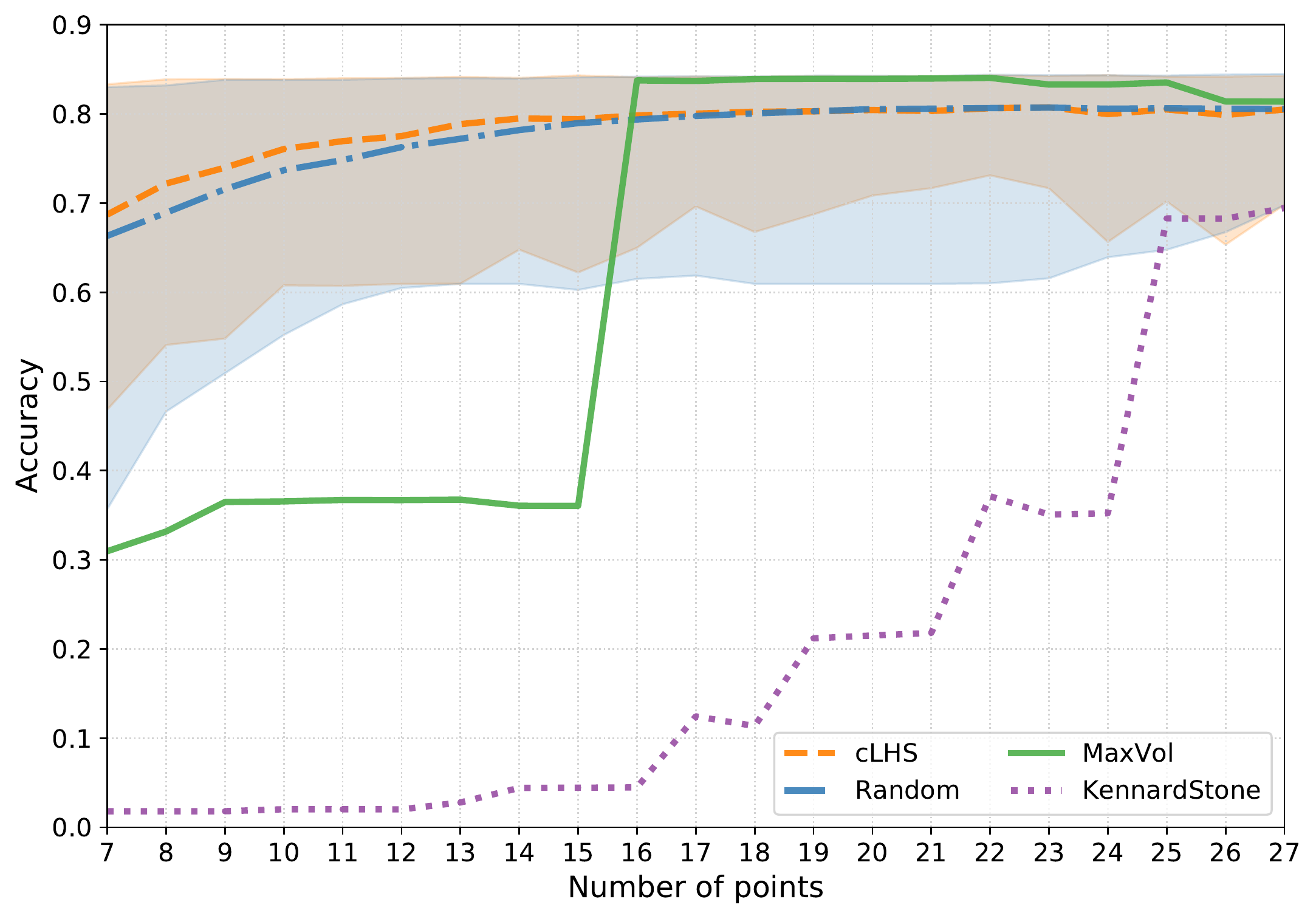}
  \caption{Comparison by balanced accuracy.}
  \label{test_sub1}
\end{subfigure}%
\begin{subfigure}{.5\linewidth}
  \centering
  \includegraphics[width=\linewidth]{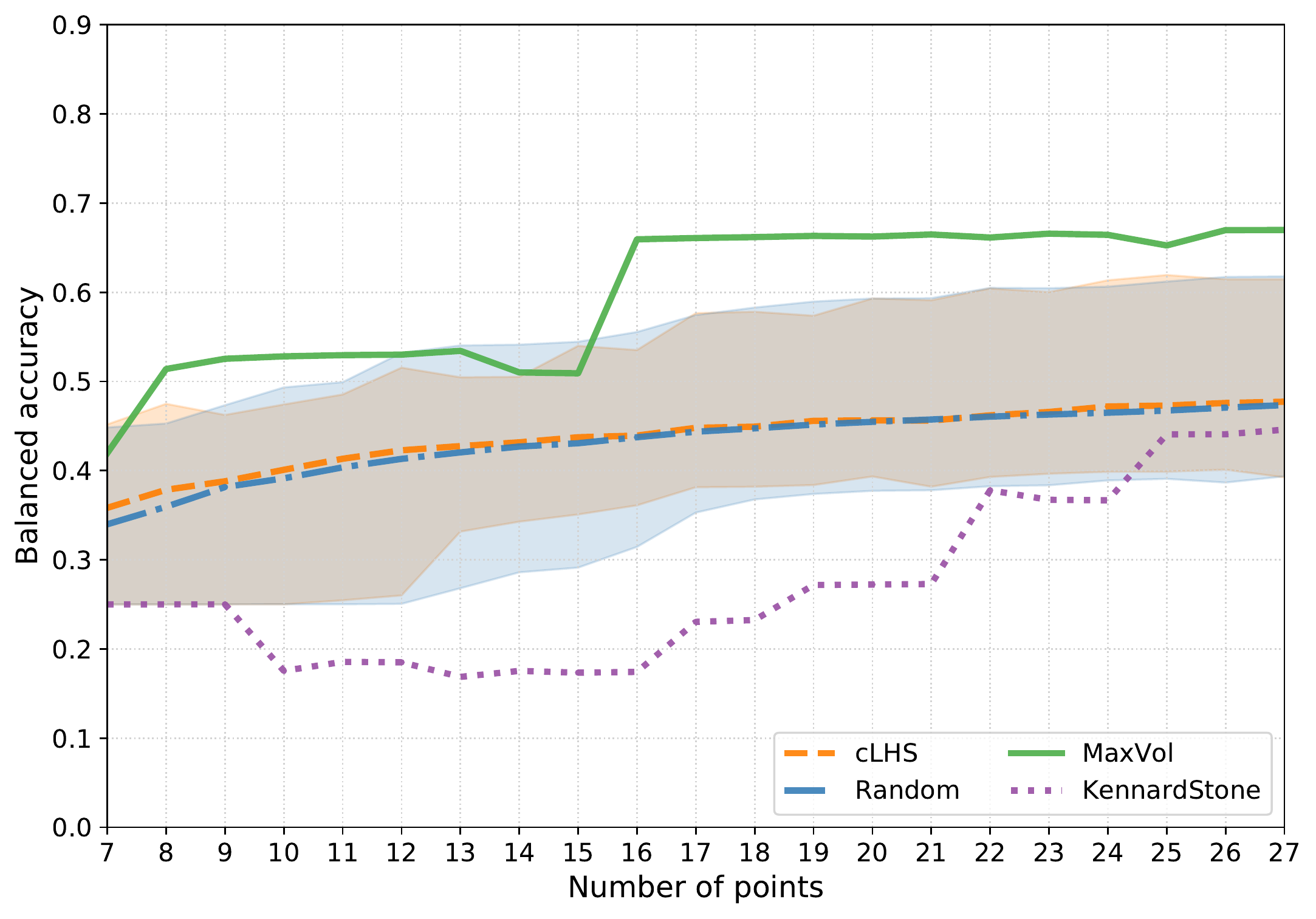}
  \caption{Comparison by plain accuracy.}
  \label{test_sub2}
\end{subfigure}
\caption{Evaluation of the {\bf{maxvol}}, cLHS, random sampling and Kennard--Stone algorithms on the "Kursk" site. For plain accuracy, the {\bf{maxvol}} algorithm reaches the upper bound of the interval of cLHS and random sampling designs and outperforms their average values starting from 16 points. For balanced accuracy, the {\bf{maxvol}} algorithm outperforms other methods for each number of points in all tested cases.}
\label{fig:test}
\end{figure}

The provided graph (\cref{fig:points_graph}) illustrates the comparison of the sampling algorithms' performance. The left graph ( \cref{test_sub1}) illustrates the algorithms' performance measured by plain accuracy score, the right one (\cref{test_sub2}) --- measured by balanced accuracy. Comparing algorithms by plain accuracy, one notices that {\bf{maxvol}} has significantly lower values with the number of points lower than 16. Starting from 16 points, it reaches the upper bound of the distribution of cLHS and random sampling designs and outperforms their average values. Still, because of an unequal distribution of classes within a dataset, a simple accuracy score does not correctly represent the algorithms' performance. 
As for balanced accuracy, {\bf{maxvol}} outperforms other methods in all tested cases. It saw a slight decrease in accuracy for 14--15 points; with this number of points the {\bf{maxvol}} shows performance close to the upper bound of the interval of cLHS and random sampling designs.

\begin{figure}[!ht]
  \center
  \includegraphics [scale=0.6]{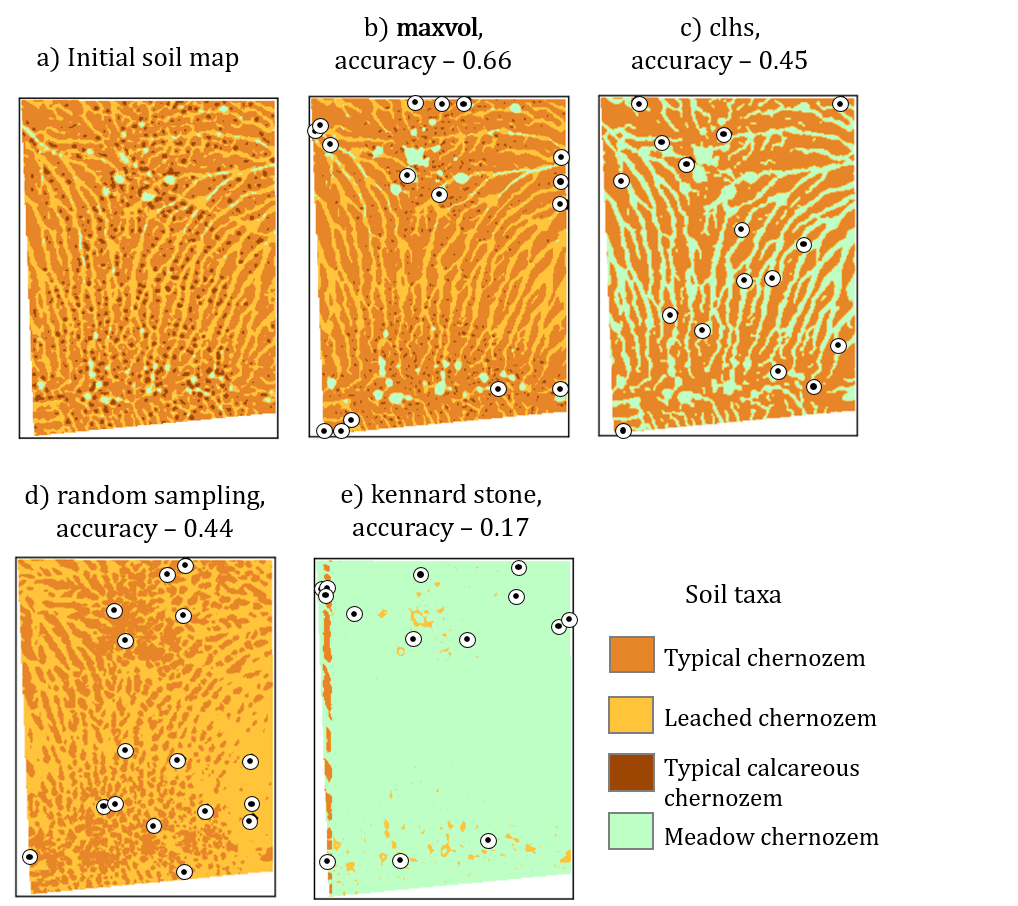}
   \caption{Soil maps obtained by performing Na\"ive Bayes prediction based on different sampling schemes. The {\bf{maxvol}} algorithm shows the best performance in capturing soil patterns.}
   \label{soil_maps} 

\end{figure}

Based on the computed results, digital soil maps were constructed according to the methodology described in \cref{eval}. 
Soil maps obtained by different sampling schemes are demonstrated in  \cref{soil_maps} (number of points = 16). The figure allows us to compare the algorithms' performances visually. Comparing the performance to the original soil map, the maxvol sampling scheme provides the best reconstruction of soil patterns. As cLHS and random sampling are randomised algorithms, one arbitrary result from 1000 was taken for each of them. Both of these algorithms capture the main pattern - soils in hollows are less carbonated (have the lower limit of carbonate occurrence) because of higher precipitation \citep{lozbenev2021incorporating}. However, particular soil taxa are predicted with high inaccuracies and, for both algorithms, typical calcareous chernozem was not taken into consideration at all as a consequence of its localisation in small spots. The spots are distinguished by the DEM, which allowed the maxvol algorithm to capture them. The Kennard–Stone algorithm shows the worst results on the test data.

\subsection{Experiments with cultivated fields}
Let us consider the results of the experiments performed on the "Kshen" and "Ozery" sites. Both experiments were conducted with maxvol adjusted with the distance constraint.  In contrast to the "Kursk" site, the data sets from cultivated fields have no information about soil cover heterogeneity. Thus, it is impossible to perform a precise estimation of maxvol’s ability to capture soil cover diﬀerences. These experiments provide an understanding of how maxvol chooses sampling locations on cultivated fields that diﬀer from virgin ones by shape, size, and elevation.

\begin{figure}[!h]

  \center
  \includegraphics[scale=1.2]{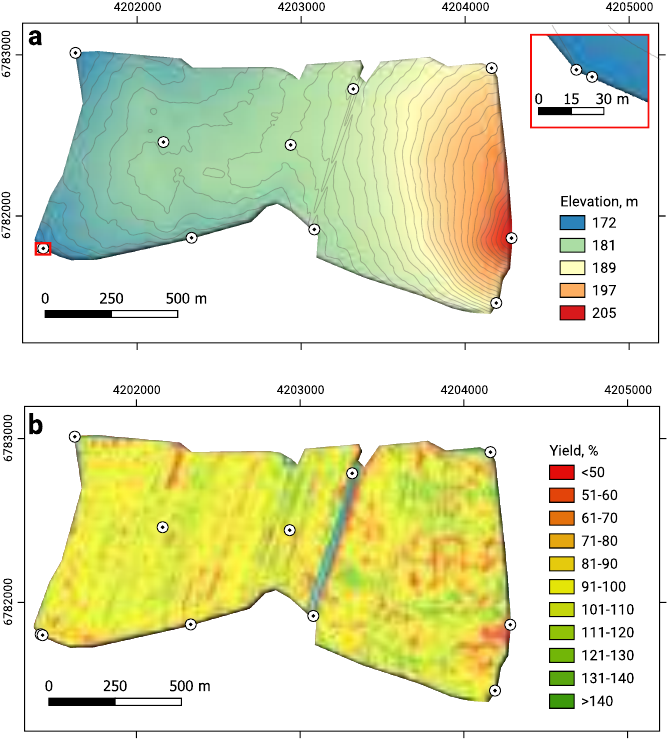}
   \caption{Sampling points chosen by the {\bf{maxvol}} on the "Kshen" site}
   \label{kshen} 
\end{figure}
 
  Fig. \ref{kshen} shows results for the "Kshen" site. In  \cref{kshen}a, the sampling scheme is demonstrated over the elevation map. The elevation of the site is rather monotonous, thus the distance constraint was fixed at a value of 0.2, which means that the minimum distance between points is approximately 5 kilometres. Sampling points are located approximately at the same intervals except for two points on the field's left corner. It can be noticed that, as in the case of the virgin field, points capture the main landforms of the site. In  \cref{kshen}b, sampling points are overlaid with the yield map of the site. The map is much more heterogeneous than the elevation on the site. Nevertheless, points are located on the most noticeable zones of high and low productivity.

\begin{figure}[!h]
  \center
  \includegraphics[scale=1.3]{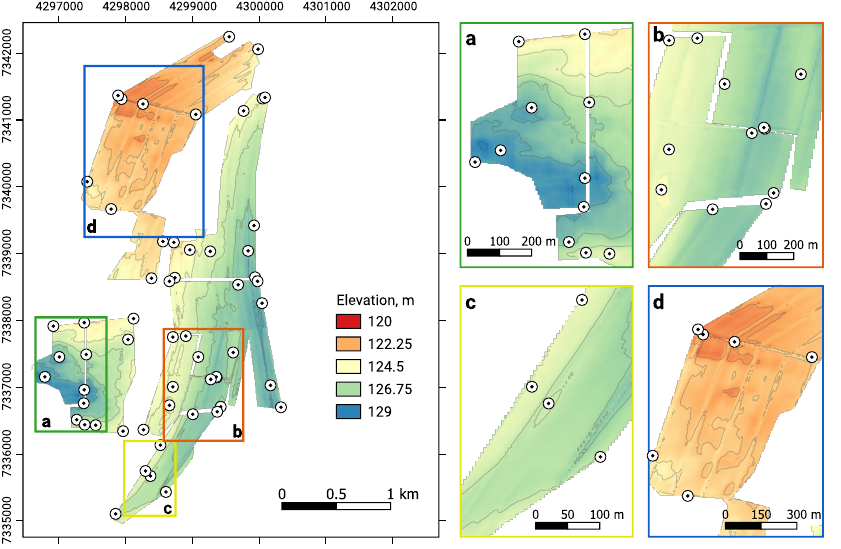}
   \caption{Sampling points chosen by {\bf{maxvol}} on the "Ozery" site}
   \label{ozery} 
\end{figure}

The results for the "Ozery" site are represented in  \cref{ozery}. On the Ozery site, every part of the field is provided by 7 or 10 points according to the size. In total, the sampling scheme contains 58 points. The site has a diverse elevation, and the success of the {\bf{maxvol}} algorithm on the site is highly influenced by the elevation pattern, as our method uses only topographical features. The distance constraint on the fields varies from about 50 meters for the smallest field with the highest ruggedness (  \cref{ozery}a) to approximately 200 meters for the biggest field ( \cref{ozery}d)

The part of the field presented on  \cref{ozery}a,b,c shows good examples of sampling schemes and {\bf{maxvol}} perfectly captures the landforms in these cases. However, there was a part of the field with unsatisfactory placement of sampling points which is shown in \cref{ozery}d. Thus, based on the results of using the algorithm, it was noted that in contours with more conspicuous elements of mesorelief, points occupy positions that most accurately describe the terrain’s heterogeneity. In contours with a more uniform mesorelief, a microrelief has a large influence on the algorithm  (eg. agricultural machinery passages), reducing the quality of the compiled sampling scheme.

\subsection{Time complexity}

In order to show that the maxvol algorithm is applicable to usual soil mapping challenges on a large scale, we provide time complexity experiments. The experiments were conducted on 1 CPU core. The {\bf{maxvol}} deals with the matrix of land-surface parameters of approximate size 64 000*5 ("Kursk" site) in
0.84 s ± 0.07 s
per loop (mean ± std. dev. of 7 runs, 1 loop each). The field on the "Kshen" site has the biggest dataset. The matrix of land-surface parameters was of the size of about 32 mln rows per 5 columns. The {\bf{maxvol}} procedure takes the following CPU times: 
6min 14s ± 2s per loop (mean ± std. dev. of 7 runs, 1 loop each).

\section{Discussion} 

The results show that the maxvol algorithm successfully produces optimal soil sampling schemes. Furthermore, it should be noted that the success of the maxvol algorithm is entirely defined by the choice of predictors. If predictors explain the majority of soil-landscape relationships in the investigated area, the maxvol algorithm can produce an optimal scheme for soil sampling. For example, in our experiment, the comparison demonstrates that maxvol outperforms popular algorithms. The maxvol performance is accounted for by the good choice of features, as the test dataset shows the high impact of microrelief on soil taxa.

The choice of predictors should be conducted concerning the main soil forming factors on the site.
Considering large-scale digital soil mapping, topography usually has the highest impact on soil patterns \citep{jana2012topographic, jana2012topography}. Thus, the usage of topographical characteristics as a predictor for sampling helps to overcome the challenge of capturing soil variability's main patterns. Nevertheless, other environmental characteristics could be used as core information to place sampling points. Various remote sensing information sources, such as satellite imagery with different spectral, temporal, radiometric, spatial resolutions, ground-penetrating radar, unmanned aerial vehicles, could provide data for improving sampling schemes. 

When comparing our results to other popular sampling methods, it can be pointed out that maxvol has several advantages. The first one is that maxvol produces only one version of the sampling scheme, so it makes the algorithm more convenient for expensive field surveys. At the same time, because of this, maxvol is less applicable to tasks which require the use of several sets of sampling points, for example, in model calibration. The second advantage is the quality of the produced sampling scheme with very few points (\cref{fig:sub2}). The {\bf{maxvol}} algorithm shows better accuracy than other methods for sampling sets consists of  7--12 points. This characteristic of the {\bf{maxvol}} algorithm allows users to start surveys with a small number of sampling points.  In general, taking into account the advantages of {\bf{maxvol}}, it can be beneficial for practical applications.

The choice of the distance constraint’s value requires more precise study. The value could be chosen based on the researcher’s intuition, auxiliary information or soil sampling regulations. One possible way is to rely on Fridland’s  \citep{fridland1974structure} concept of soil mantle structure. According to this concept, soil cover consists of elementary soil areals, the size of which varies depending on the influence of soil-forming factors. If the typical size of elementary soil areals for the investigated area is known, it could be the reference value for the smallest distance between sampling points.

This research presents the maxvol algorithm’s applicability in large-scale digital soil mapping, but the maxvol algorithm can equally be applied to creating soil maps on smaller scales. The usage of maxvol for various spatial sampling cases should be considered in further research.

\section{Conclusion}
A new sampling algorithm based on an optimal-design approach was presented. The sampling design is then tested in three real cases diﬀering in field data availability and compared to the popular sampling schemes. It is shown that maxvol can outperform popular sampling methods in soil type prediction based on land-surface parameters, and deals with massive datasets in a reasonable time.

\section{Acknowledgements}
The authors wish to thank N. Lozbenev and D. Kozlov for the materials provided on the "Kursk" site. The authors are also grateful to M. Pukalchik, P. Tregubova, A. Nikitin, M. Gasanov who participated in data collection. The research was supported by the Ministry of Science and Higher Education of the Russian Federation Agreement 075-10-2020-091 (grant 14.756.31.0001).

\bibliography{mybibfile}

\begin{thebibliography}{10}
\expandafter\ifx\csname url\endcsname\relax
  \def\url#1{\texttt{#1}}\fi
\expandafter\ifx\csname urlprefix\endcsname\relax\def\urlprefix{URL }\fi
\expandafter\ifx\csname href\endcsname\relax
  \def\href#1#2{#2} \def\path#1{#1}\fi

\bibitem{hengl2003soil}
T.~Hengl, D.~G. Rossiter, A.~Stein, Soil sampling strategies for spatial
  prediction by correlation with auxiliary maps, Soil Research 41~(8) (2003)
  1403--1422.

\bibitem{brus2007optimization}
D.~J. Brus, G.~B. Heuvelink, Optimization of sample patterns for universal
  kriging of environmental variables, Geoderma 138~(1-2) (2007) 86--95.

\bibitem{zhu2008purposive}
A.~X. Zhu, L.~Yang, B.~Li, C.~Qin, E.~English, J.~E. Burt, C.~Zhou, Purposive
  sampling for digital soil mapping for areas with limited data, in: Digital
  soil mapping with limited data, Springer, 2008, pp. 233--245.

\bibitem{fitzgerald2006directed}
G.~J. Fitzgerald, S.~M. Lesch, E.~M. Barnes, W.~E. Luckett, Directed sampling
  using remote sensing with a response surface sampling design for
  site-specific agriculture, Computers and electronics in agriculture 53~(2)
  (2006) 98--112.

\bibitem{clifford2014pragmatic}
D.~Clifford, J.~E. Payne, M.~Pringle, R.~Searle, N.~Butler, Pragmatic soil
  survey design using flexible latin hypercube sampling, Computers \&
  Geosciences 67 (2014) 62--68.

\bibitem{liess2015sampling}
M.~Lie{\ss}, Sampling for regression-based digital soil mapping: closing the
  gap between statistical desires and operational applicability, Spatial
  Statistics 13 (2015) 106--122.

\bibitem{musafer2016optimal}
G.~N. Musafer, M.~Thompson, Optimal adaptive sequential spatial sampling of
  soil using pair-copulas, Geoderma 271 (2016) 124--133.

\bibitem{nawar2018optimal}
S.~Nawar, A.~M. Mouazen, Optimal sample selection for measurement of soil
  organic carbon using on-line vis-nir spectroscopy, Computers and Electronics
  in Agriculture 151 (2018) 469--477.

\bibitem{nketia2019new}
K.~A. Nketia, S.~B. Asabere, S.~Erasmi, D.~Sauer, A new method for selecting
  sites for soil sampling, coupling global weighted principal component
  analysis and a cost-constrained conditioned latin hypercube algorithm,
  MethodsX 6 (2019) 284--299.

\bibitem{aktas2019landslide}
H.~Aktas, B.~T. San, Landslide susceptibility mapping using an automatic
  sampling algorithm based on two level random sampling, Computers \&
  Geosciences 133 (2019) 104329.

\bibitem{wadoux2019sampling}
A.~M.-C. Wadoux, D.~J. Brus, G.~B. Heuvelink, Sampling design optimization for
  soil mapping with random forest, Geoderma 355 (2019) 113913.

\bibitem{castaldi2019sampling}
F.~Castaldi, S.~Chabrillat, B.~Van~Wesemael, Sampling strategies for soil
  property mapping using multispectral sentinel-2 and hyperspectral enmap
  satellite data, Remote Sensing 11~(3) (2019) 309.

\bibitem{liess2020interface}
M.~Lie{\ss}, At the interface between domain knowledge and statistical sampling
  theory: Conditional distribution based sampling for environmental survey
  (codibas), Catena 187 (2020) 104423.

\bibitem{ma2020comparison}
T.~Ma, D.~J. Brus, A.-X. Zhu, L.~Zhang, T.~Scholten, Comparison of conditioned
  latin hypercube and feature space coverage sampling for predicting soil
  classes using simulation from soil maps, Geoderma 370 (2020) 114366.

\bibitem{yang2020evaluation}
L.~Yang, X.~Li, J.~Shi, F.~Shen, F.~Qi, B.~Gao, Z.~Chen, A.-X. Zhu, C.~Zhou,
  Evaluation of conditioned latin hypercube sampling for soil mapping based on
  a machine learning method, Geoderma 369 (2020) 114337.

\bibitem{minasny2006conditioned}
B.~Minasny, A.~B. McBratney, A conditioned latin hypercube method for sampling
  in the presence of ancillary information, Computers \& geosciences 32~(9)
  (2006) 1378--1388.

\bibitem{de2006sampling}
J.~De~Gruijter, D.~J. Brus, M.~F. Bierkens, M.~Knotters, Sampling for natural
  resource monitoring, Springer Science \& Business Media, 2006.

\bibitem{brus1997random}
D.~Brus, J.~De~Gruijter, Random sampling or geostatistical modelling? choosing
  between design-based and model-based sampling strategies for soil (with
  discussion), Geoderma 80~(1-2) (1997) 1--44.

\bibitem{brus2012hybrid}
D.~Brus, J.~De~Gruijter, A hybrid design-based and model-based sampling
  approach to estimate the temporal trend of spatial means, Geoderma 173 (2012)
  241--248.

\bibitem{OptDesign}
V.~Fedorov, Theory Of Optimal Experiments, Probability and Mathematical
  Statistics, Elsevier Science, 1972.

\bibitem{natrella2010nist}
M.~Natrella, et~al., Nist/sematech e-handbook of statistical methods,
  Nist/Sematech 49.

\bibitem{jana2012topographic}
R.~B. Jana, B.~P. Mohanty, On topographic controls of soil hydraulic parameter
  scaling at hillslope scales, Water Resources Research 48~(2).

\bibitem{jana2012topography}
R.~B. Jana, B.~P. Mohanty, A topography-based scaling algorithm for soil
  hydraulic parameters at hillslope scales: Field testing, Water Resources
  Research 48~(2).

\bibitem{kozlov2017soil}
D.~Kozlov, E.~Levchenko, N.~Lozbenev, Soil combinations as an object of dsm: a
  case study in chernozems area of the russian plain, in: Proceedings of the
  Global Soil Map 2017 Conference, 2017, pp. 81--88.

\bibitem{lozbenev2021incorporating}
N.~Lozbenev, A.~Yurova, M.~Smirnova, D.~Kozlov, Incorporating process-based
  modeling into digital soil mapping: A case study in the virgin steppe of the
  central russian upland, Geoderma 383 (2021) 114733.

\bibitem{kennard1969computer}
R.~W. Kennard, L.~A. Stone, Computer aided design of experiments, Technometrics
  11~(1) (1969) 137--148.

\bibitem{liu2011wisdom}
N.~N. Liu, X.~Meng, C.~Liu, Q.~Yang, Wisdom of the better few: cold start
  recommendation via representative based rating elicitation, in: Proceedings
  of the fifth ACM conference on Recommender systems, 2011, pp. 37--44.

\bibitem{wang2010global}
B.~H. Wang, H.~T. Hui, M.~S. Leong, Global and fast receiver antenna selection
  for mimo systems, IEEE Transactions on Communications 58~(9) (2010)
  2505--2510.

\bibitem{Ryzhakov2018}
G.~Ryzhakov, I.~Oseledets, Function approximation using gradient information
  with application to parametric and stochastic differential equations, arXiv
  preprint arXiv:1802.01542.

\bibitem{goreinov2010find}
S.~A. Goreinov, I.~V. Oseledets, D.~V. Savostyanov, E.~E. Tyrtyshnikov, N.~L.
  Zamarashkin, How to find a good submatrix, in: Matrix Methods: Theory,
  Algorithms And Applications: Dedicated to the Memory of Gene Golub, World
  Scientific, 2010, pp. 247--256.

\bibitem{de1995d}
P.~F. de~Aguiar, B.~Bourguignon, M.~Khots, D.~Massart, R.~Phan-Than-Luu,
  D-optimal designs, Chemometrics and intelligent laboratory systems 30~(2)
  (1995) 199--210.

\bibitem{mikhalev2018rectangular}
A.~Mikhalev, I.~V. Oseledets, Rectangular maximum-volume submatrices and their
  applications, Linear Algebra and its Applications 538 (2018) 187--211.

\bibitem{egorov1987classification}
V.~V. Egorov, V.~Fridland, E.~Ivanova, N.~Rozov, V.~Nosin, T.~Friev, et~al.,
  Classification and diagnostics of soils of the ussr., Classification and
  diagnostics of soils of the USSR.

\bibitem{karklins2015world}
A.~Karklins, et~al., World reference base for soil resources-the new edition
  2014, in: Zin{\=a}tniski praktisk{\=a} konference: L{\=\i}dzsvarota
  lauksaimniec{\=\i}ba,, Jelgava (Latvia), 19--20 Feb 2015, Latvijas
  Lauksaimniec{\=\i}bas universit{\=a}te, 2015.

\bibitem{lozbenev2019digital}
N.~Lozbenev, M.~Smirnova, M.~Bocharnikov, D.~Kozlov, Digital mapping of habitat
  for plant communities based on soil functions: a case study in the virgin
  forest-steppe of russia, Soil Systems 3~(1) (2019) 19.

\bibitem{saga}
S.~U.~G. Association, Saga—system for automated geoscientific analysis,
  \url{http://www.saga-gis.org}, last accessed 2020-01-12 (2020).

\bibitem{GRASS_GIS_software}
G.~D. Team, Geographic resources analysis support system (grass) software,
  version 7.8. open source geospatial foundation,
  \url{https://grass.osgeo.org}, last accessed 2020-01-12 (2020).

\bibitem{einsmann1999nutrient}
J.~C. Einsmann, R.~H. Jones, M.~Pu, A.~J. Mitchell, Nutrient foraging traits in
  10 co-occurring plant species of contrasting life forms, Journal of Ecology
  87~(4) (1999) 609--619.

\bibitem{hutchings2003toward}
M.~J. Hutchings, E.~A. John, D.~K. Wijesinghe, Toward understanding the
  consequences of soil heterogeneity for plant populations and communities,
  Ecology 84~(9) (2003) 2322--2334.

\bibitem{patzold2008soil}
S.~Patzold, F.~M. Mertens, L.~Bornemann, B.~Koleczek, J.~Franke, H.~Feilhauer,
  G.~Welp, Soil heterogeneity at the field scale: a challenge for precision
  crop protection, Precision Agriculture 9~(6) (2008) 367--390.

\bibitem{fridland1974structure}
V.~Fridland, Structure of the soil mantle, Geoderma 12~(1-2) (1974) 35--41.

\end{thebibliography}

\end{document}